%% file: main.tex
\documentclass[conference]{IEEEtran}
\IEEEoverridecommandlockouts
\usepackage{cite}
\usepackage{amsmath,amssymb,amsfonts}
\usepackage{algorithmic}
\usepackage{graphicx}
\usepackage{textcomp}
\usepackage{enumitem}
\usepackage{xcolor}
\usepackage{booktabs}
\usepackage{multirow}
\usepackage{tabularx}
\usepackage{siunitx}
\usepackage{url}
\usepackage[a4paper, total={184mm,239mm}]{geometry}
\def\BibTeX{{\rm B\kern-.05em{\sc i\kern-.025em b}\kern-.08em
    T\kern-.1667em\lower.7ex\hbox{E}\kern-.125emX}}
\begin{document}

\title{TurboSAT: Gradient-Guided Boolean Satisfiability \\Accelerated on GPU-CPU Hybrid System\vspace{-1.7pt}}

\author{\IEEEauthorblockN{Steve Dai\textsuperscript{*}, Cunxi Yu, Kalyan Krishnamani, Brucek Khailany\vspace{-1.1pt}}
\IEEEauthorblockA{
\textit{NVIDIA}\\
\textsuperscript{*}sdai@nvidia.com}
% \and
}

\maketitle
\thispagestyle{plain}
\pagestyle{plain}

\input{0_abstract}

%\begin{IEEEkeywords}
%component, formatting, style, styling, insert
%\end{IEEEkeywords}

\input{1_introduction}
\input{2_preliminaries}
\input{3_formulation}
\input{4_system}
\input{5_experiments}
\input{6_conclusions}
\clearpage
\bibliographystyle{ieeetr}
\bibliography{references.bib}

\end{document}

%% file: 0_abstract.tex
\begin{abstract}
While accelerated computing has transformed many domains of computing, its impact on logical reasoning, specifically Boolean satisfiability (SAT), remains limited.
State-of-the-art SAT solvers rely heavily on inherently sequential conflict-driven search algorithms that offer powerful heuristics but limit the amount of parallelism that could otherwise enable significantly more scalable SAT solving.
Inspired by neural network training, we formulate the SAT problem as a binarized matrix-matrix multiplication layer that could be optimized using a differentiable objective function.
Enabled by this encoding, we combine the strengths of parallel differentiable optimization and sequential search to accelerate SAT on a hybrid GPU-CPU system.
%In this system, the GPUs leverage parallel neural network training to rapidly evaluate SAT clauses and use gradients to stochastically explore the solution space and optimize variable assignments.
In this system, the GPUs leverage parallel differentiable solving to rapidly evaluate SAT clauses and use gradients to stochastically explore the solution space and optimize variable assignments.
Promising partial assignments generated by the GPUs are post-processed on many CPU threads which exploit conflict-driven sequential search to further traverse the solution subspaces and identify complete assignments.
Prototyping the hybrid solver on an NVIDIA DGX GB200 node, our solver achieves runtime speedups up to over 200x when compared to a state-of-the-art CPU-based solver on public satisfiable benchmark problems from the SAT Competition.
\end{abstract}

%% file: 1_introduction.tex
\section{Introduction}
\label{sec:intro}

\begin{figure}[t]
    \centering
    \includegraphics[trim=0pt 363pt 486pt 0pt, clip, width=0.8\linewidth]{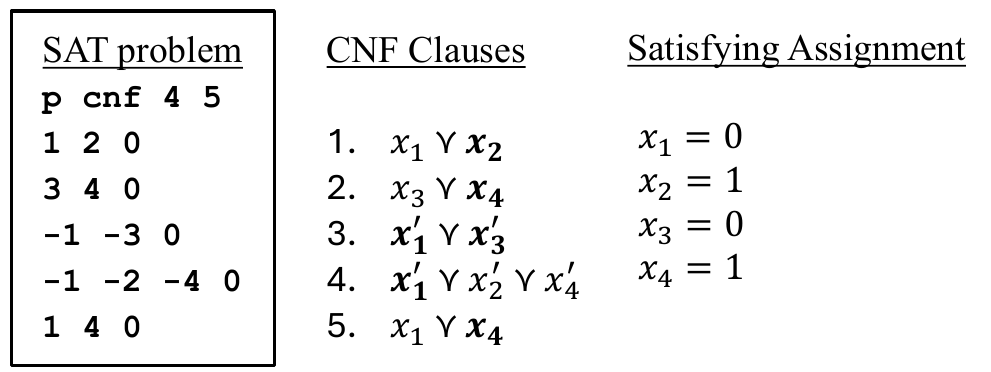}
    \caption{
    \textbf{SAT problem in CNF with 4 variables and 5 clauses} -- 
%    The satisfying assignment satisfies the bold literals in each clause. 
    The problem is satisfiable because each clause has at least one satisfied (bold) literal.
    }
    \label{fig:cnf}
\end{figure}

Boolean satisfiability (SAT) is a fundamental NP-complete problem of determining whether there exists an assignment of truth values to variables that satisfies a given Boolean formula~\cite{HandbookSAT2021}.
As shown in Figure~\ref{fig:cnf}, a SAT formula is typically expressed in Conjunctive Normal Form (CNF) consisting of a conjunction of clauses where each clause is a disjunction of literals.
Each literal represents a variable or its negation, and each variable can be assigned a truth value of either 1 (True) or 0 (False).
A SAT problem is considered satisfiable if there exists at least one assignment of the variables such that every clause contains at least one true literal.
Otherwise, the problem is deemed unsatisfiable (UNSAT).
Figure~\ref{fig:cnf} shows a satisfiable problem along with a satisfying assignment.
% As a cornerstone of modern logical reasoning, SAT serves as an integral component of computer-aided design (CAD), enabling critical formal verification tasks such as bounded model checking and equivalence checking. 
% In addition, SAT plays a critical role in many challenging combinatorial problems.
As a cornerstone of modern logical reasoning, SAT serves as an integral component of computer-aided design (CAD) and plays a critical role in challenging combinatorial problems.

Modern SAT solvers typically employ a systematic search algorithm known as Conflict-Driven Clause Learning (CDCL) which explores variable assignments through an iterative process of decision, propagation, conflict learning, and backtracking~\cite{MarquesSilva1996GRASP}.
CDCL’s effectiveness stems from its ability to prune future search spaces using learned conflict clauses and to backtrack non-chronologically to specific decision level responsible for a conflict, skipping redundant exploration.
Depending on the particular solver, CDCL is often complemented with various specialized heuristics to further improve efficiency.
For example, \mbox{MiniSat} introduces phase saving to remember previously assigned values~\cite{Een2003ExtensibleSAT}.
%Chaff implements Variable State Independent Decaying Sum (VSIDS) to record variable scores and prioritize variables involved in recent conflicts during branching decisions~\cite{Moskewicz2001Chaff}.
\mbox{CaDiCaL} offers clause subsumption to remove clauses containing redundant information~\cite{BiereFallerFazekasFleuryFroleyksPollitt-CAV24}.
%CDCL-based solvers are integrated within CAD tools to tackle industrial-grade problems~\cite{Ganesh2021UnreasonableSAT}.
CDCL-based solvers are often integrated within CAD tools to tackle industrial-grade problems in formal verification, automatic test generation, and other chip design tasks~\cite{Ganesh2021UnreasonableSAT}.

Despite its success, CDCL relies on an inherently sequential search algorithm, which limits the potential for parallelism.
Prior efforts to parallelize SAT solving based on CDCL include performing divide-and-conquer on the decision tree~\cite{LeFrioux2019DivideAndConquer} or employing a portfolio of solvers with different policies and heuristics~\cite{Hamadi2009ManySAT}.
In addition, several SAT solvers have been developed to leverage the parallel processing power of GPUs. 
For example, ParaFROST accelerates inprocessing on GPU with parallel variable elimination, eager redundancy elimination, and a GPU-based garbage collector~\cite{osama2021sat}.
Fujii and Fujimoto parallelizes Boolean Constraint Propagation (BCP) on GPU with clause partitioning, allowing different groups of clauses to be processed in parallel during the propagation phase~\cite{Fujii2012GPUBCP}.
% In contrast to CDCL, stochastic local search (SLS) algorithms such as WalkSAT are intrinsically parallelizable, allowing many assignments to be explored independently in parallel~\cite{selman1994walksat}. 
% SLS algorithms start with a set of randomly initialized complete assignments, track variables in unsatisfied clauses, and iteratively update selected variables in those clauses using a score-based probabilistic heuristic for variable-flipping.
% While highly parallelizable, SLS methods tend to excel only on uniform random synthetic SAT instances that are not overly constrained.
% Overall, they do not learn conflict clauses which would facilitate efficiently reducing the search space and improving convergence.

Driven by the demands of artificial intelligence and high-performance computing, heterogeneous architectures that tightly integrate GPUs, CPUs, memory, and interconnects are emerging as the go-to systems for accelerated computing.
NVIDIA’s Grace Blackwell Superchip exemplifies this trend by combining the Blackwell GPU with the many-core Grace CPU through a unified memory model and high-speed NVLink interconnects~\cite{nvidia_grace_blackwell_document}.
% Notably, Tensor Cores are optimized for large-scale matrix operations typical in deep learning (DL) workloads and support low-precision data formats such as FP8 to dramatically enhance computational throughput and reduce memory burden.
% On the other hand, many-core CPUs excel at executing an abundance of sequential tasks simultaneously across many threads, ensuring versatility for diverse and dynamic workload demands.
Notably, GPUs are optimized for large-scale parallel matrix operations typical in deep learning (DL) workloads, while many-core CPUs excel at executing an abundance of sequential tasks simultaneously across many threads.
We henceforth refer to a GPU-CPU system as a hybrid system in the rest of the paper. 
%Care should be taken not to confuse this with hybrid systems of control theory.
%The growing potential of such hybrid systems reflects the hybrid nature of modern workloads and underscores the need to redesign or evolve traditional algorithms to take full advantage of these accelerated architectures.
%Achieving meaningful acceleration requires the algorithms to effectively balance exploration versus exploitation and parallel versus sequential processing. 
%In this context, hybrid systems present exciting opportunities to re-imagine SAT solving beyond merely traditional CDCL and SLS approaches, enabling synergy between the exploratory power of parallel optimization on GPU and the exploitation power of sequential search on CPU.
%The growing potentials of such hybrid systems present exciting opportunities to re-imagine SAT solving beyond traditional CDCL, enabling synergy between the exploratory power of parallel optimization on GPU and the exploitation power of sequential search on CPU.
These hybrid systems present opportunity to re-imagine SAT solving beyond traditional CDCL, combining the exploratory power of parallel optimization on GPU and the exploitation power of sequential search on CPU.

%To address these opportunities, we present a novel SAT solver with a hybrid algorithm that harnesses the GPU-CPU synergy to efficiently tackle SAT solving in two distinct phases.
%Our algorithm strategically combines GPU-enabled parallel differentiable optimization and CPU-friendly sequential conflict-driven search to dramatically accelerate SAT solving using an exploration-exploitation approach.
To address these opportunities, we present a novel SAT algorithm that harnesses the GPU-CPU synergy to efficiently tackle SAT solving in two distinct phases:  GPU-enabled parallel differentiable optimization and CPU-friendly conflict driven search.
This algorithm dramatically accelerates SAT solving using an exploration-exploitation approach.
Specifically, our key contributions are as follows:
\begin{itemize}[leftmargin=10pt]
    \item \textit{Differentiable Formulation:} We formulate the SAT problem as a binary matrix-matrix multiplication layer paired with a differentiable objective function for SAT solving. 
    This formulation enables gradient-guided optimization of variable assignments to maximize satisfiability.
    \item \textit{Massively Parallel Exploration:} We utilize differentiable solving on the GPU to iteratively evaluate SAT clauses and optimize variable assignments with massive parallelism. 
    This allows the solver to rapidly explore the large solution space and quickly identify promising solution subspaces.
    \item  \textit{Conflict-Driven Exploitation:} 
    We post-process promising partial assignments identified by the GPU with many parallel CPU threads.
    %We offload promising partial assignments from the GPU to many CPU threads. 
    Each thread exploits conflict-driven learning and sequential search to further traverse solution subspaces and identify complete assignments efficiently.
    \item \textit{Implementation:} We prototype our solver on a DGX GB200 GPU-CPU system using one GPU instance and 100 CPU threads. 
    On satisfiable benchmarks from the 2024 SAT Competition, our hybrid solver achieves runtime speedup up to over 200x over a state-of-the-art CPU-only solver. %with the same \mbox{MiniSat} solver.
\end{itemize}
% The remainder of this paper is organized as follows:
% Section~\ref{sec:preliminaries} introduces the preliminaries of accelerated SAT solving.
% Section~\ref{sec:formulation} presents our differentiable SAT formulation and model optimization methodology.
% Section~\ref{sec:system} describes the hybrid system for solving SAT using GPU and CPU.
% Section~\ref{sec:experiments} details our solver prototype and reports experimental results.
% Section~\ref{sec:conclusions} summarizes our contributions and discusses potential future directions.

%% file: 2_preliminaries.tex
\begin{figure*}[t]
    \centering
    \includegraphics[trim=0pt 280pt 208pt 6pt, clip, width=0.95\linewidth]{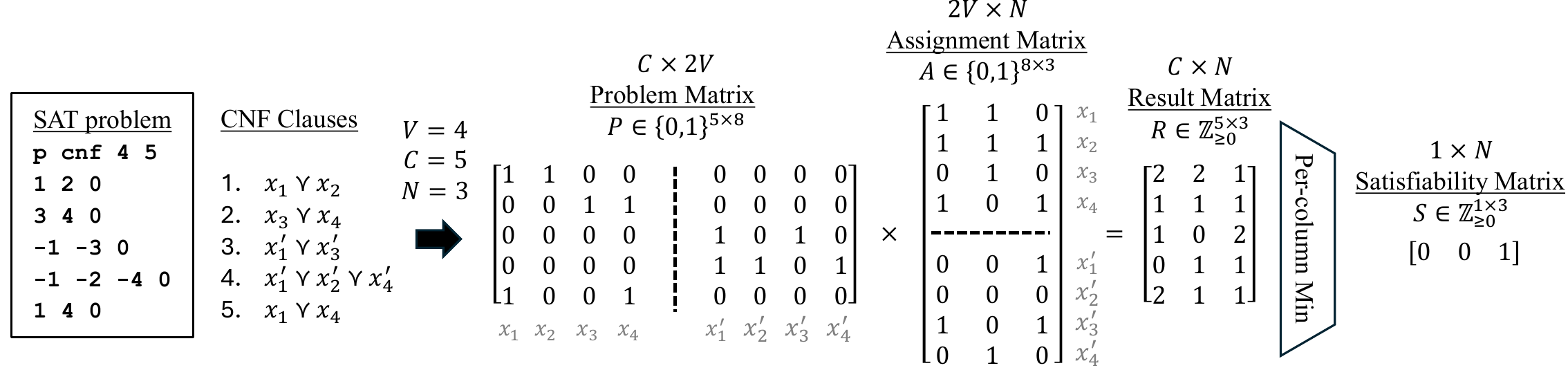}
    \caption{\textbf{
    Binary matrix multiplication encoding for our differentiable SAT formulation} -- 
    Each row of the Problem Matrix encodes one corresponding CNF clause. 
    Each column of the Assignment Matrix represents one candidate assignment being attempted.
    Each element in the Result Matrix indicates the number of literals that each candidate assignment satisfies in each clause.
    The Satisfiability Matrix indicates whether each assignment is satisfying all clauses.
    }
    \label{fig:formulation}
\end{figure*}

\section{Related Work}
\label{sec:preliminaries}

Mirroring the broader trend of machine learning for CAD, neural networks have emerged as potential candidates for augmenting or even replacing traditional SAT algorithms~\cite{guo2023machine}.
In this context, recent work has introduced both supervised and unsupervised learning techniques as new paradigms for SAT solving.
These techniques leverage specialized neural network architectures to represent SAT instances by encoding variables, clauses, and their constraints into continuous embeddings.
% At the same time, significant efforts have been directed towards improving the efficiency of neural network execution through quantization techniques.
% At the extreme end of this spectrum, BNNs represent both weights and activations using single-bit values.
% In the following sections, we survey learning-based SAT solving techniques across supervised and unsupervised paradigms and introduce quantization and BNNs to highlight their relevance in logical reasoning.

\subsection{Supervised Learning for SAT}
Supervised learning techniques for SAT solving involve training neural networks on labeled SAT instances to predict solutions or guide solver heuristics.
NeuroSAT proposes an end-to-end approach that leverages Graph Neural Network models to encode the relationship among the variables and treat SAT solving as a binary classification problem~\cite{Selsam2019NeuroSAT}. 
Satisfying assignments are decoded from the learned variable embeddings.
SATFormer extends this paradigm with hierarchical Transformer-based architectures to model clause correlations, enabling both satisfiability prediction and identification of unsatisfiable sub-problems, also known as UNSAT cores~\cite{shi2023satformer}.
While these end-to-end methods demonstrate some promise, they require large labeled datasets and remain less efficient and accurate than state-of-the-art CDCL solvers in practice.
Other supervised learning based approaches focus on enhancing specific heuristics used by traditional SAT solvers.
For example, NeuroCore predicts variables involved in the UNSAT core to influence variable branching decisions~\cite{Selsam2019NeuroCore}. 
NeuroBack predicts values of variables that tend to appear in satisfying assignments~\cite{Wang2024NeuroBack}.
Despite their emphasis on specific heuristics, these techniques depend on extensive training data and can struggle to generalize to unseen problem distributions.
%Our approach distinguishes from these supervised learning techniques because it requires no training data other than the SAT problem itself which is available by default.

\subsection{Unsupervised Learning for SAT}
Unsupervised learning techniques for SAT solving eliminate the need for labeled data or pre-solved instances by reformulating SAT problems into optimization tasks that directly learn the SAT solutions without prior model training.
FourierSAT transforms SAT solving into continuous optimization via Walsh-Fourier expansions, converting Boolean constraints into multilinear polynomials and guiding search through gradient-driven local improvements instead of combinatorial enumeration~\cite{kyrillidis2020fouriersat,Cen2023FastFourierSAT}.
However, this method provides incremental improvement over existing methods.
%FastFourierSAT accelerates this approach with GPU-parallelized continuous optimization, leveraging Fast Fourier Transform to compute elementary symmetric polynomials, a computationally intensive step in gradient computation~\cite{Cen2023FastFourierSAT}.
%However, FastFourierSAT requires high-precision arithmetic to mitigate numerical instability and cannot run efficiently using modern GPU features like Tensor Cores that are optimized for lower-precision operations.
DiffSAT introduces a differentiable MaxSAT layer that transforms SAT solving into a gradient-based minimization task~\cite{Zhang2024DiffSAT}. 
It initializes assignments via semidefinite relaxation and iteratively refines them using a clause-violation loss function.
%In each iteration, DiffSAT analyzes the gradients to predict the variable that contributes the most to unsatisfiability and uses gradient descent to update the assignment of that particular variable with a certain probability.
However, the algorithm's single-variable update strategy limits parallelism, hindering time to convergence and scalability on GPU architectures.

%% file: 3_formulation.tex
\section{Differentiable Formulation}
\label{sec:formulation}

Following the unsupervised learning approach, we propose a novel SAT-solving framework that combines matrix-based problem encoding inspired by GPU4SAT~\cite{jaillet2008gpu4sat} with differentiable optimization. %inspired by BNN.
Unlike existing methods, our approach focuses on leveraging GPU-accelerated General Matrix Multiplication (GEMM) to evaluate SAT problems with efficiency and massive parallelism, essentially transforming combinatorial search into a high-throughput optimization task. 
For the rest of this section, let's consider a SAT problem represented in the CNF format that contains $V$ variables and $C$ clauses. 
For convenience, we will illustrate our formulation using a small SAT problem with 4 variables ($V=4$) and 5 clauses ($C=5$) in Figure~\ref{fig:formulation}, same as the problem in Section~\ref{sec:intro} and Figure~\ref{fig:cnf}.

\subsection{Binary Matrix Encoding}
\label{sec:encoding}
We encode the SAT problem into a matrix-matrix multiplication $R=PA$ in which the Problem Matrix $P\in \{0,1\}^{C \times 2V}$ represents the clauses and literals and the Assignment Matrix $A\in \{0,1\}^{2V \times N}$ represents the variable assignments. 
The Result Matrix $R\in \mathbb{Z}_{\geq 0}^{C \times N}$ indicates the satisfiability of each clause in the problem.
$N$ is the runtime parameter that denotes the number of SAT assignments that we attempt.
This encoding allows for the parallel evaluation of all $N$ assignments on all $C$ clauses using GPU hardware. %especially the Tensor Core units. 
%By further constraining the problem matrix $P$ and assignment matrix $A$ to binary number formats, we can efficiently perform these operations on low-precision Tensor Cores, if available.

\subsubsection{Problem Matrix} 
We encode the SAT problem in CNF format into a Problem Matrix $P$ with $C$ rows and $2V$ columns.
Each row represents a single clause, and each column represents a literal, meaning a single variable or its negation.
%To ensure compatibility with anticipated binary GEMM hardware, 
We allow only 0's and 1's in the matrix, which requires allocating two separate columns in $P$ for each variable in the SAT problem.
%If an additional representation such as $-1$ is available, the matrix can be more compact with only $V$ columns. 
In our encoding, a value of 1 in $P$ indicates that the corresponding literal is present in the clause, while a 0 indicates that the corresponding literal is absent in the clause. 
As illustrated in Figure~\ref{fig:formulation}, the second clause maps to 1's in the $x_3$ and $x_4$ columns in the second row of $P$.
$P$ is analogous to the input activation tensor in a DL layer.

\subsubsection{Assignment Matrix} 
We encode the $N$ assignments being attempted by the solver into an Assignment Matrix $A$ with $2V$ rows and $N$ columns, where $N$ can be increased to attempt more assignments and explore a larger search space. 
In $A$, 0 indicates that the corresponding literal is assigned False, and 1 indicates that the corresponding literal is assigned True. 
In Figure~\ref{fig:formulation}, for example, the second assignment assigns $x_1$, $x_2$, and $x_3$ to True while assigning $x_4$ to False. 
We consider only complete assignments in $A$ where every variable is assigned either True or False. 
Therefore, a literal always has an assignment which is opposite to that of its negated counterpart. 
For example, in the second assignment, because $x_1$ is True, $x_1'$ must be False. 
Same goes for all other variables and assignments.
% However, for an incomplete assignment, a literal and its negated counterpart can be both 0's or 1's at the same time. 
% If both are 1's, the variable is considered a don't care. 
% As a result, any clause containing that variable is always satisfied. 
% If both are 0's, the variable cannot contribute to satisfying any clauses. 
% Any satisfied clauses must be satisfied by some other variables.
%For now, we consider only complete assignments in $A$.
$A$ is analogous to the weight tensor in a DL layer.

\subsubsection{Result Matrix} 
Multiplying the Problem Matrix $P$ into the Assignment Matrix $A$ produces the Result Matrix $R$ with $C$ rows and $N$ columns. 
Each column represents the result of the corresponding assignment. 
Each row in an assignment denotes the number of literals in the corresponding clause that the assignment is able to satisfy (i.e., evaluate to True). 
In Figure~\ref{fig:formulation}, Row 3 Column 2 of the Result Matrix is 0 because the second assignment assigns neither $x_1$ nor $x_3$ of the third clause to False. 
%Similarly, Row 4 Column 1 of the Result Matrix is 0 because the first assignment assigns neither $x_1$, $x_2$, nor $x_4$ of the fourth clause to False. 
On the other hand, Row 3 Column 3 of the Result Matrix is 2 because the third assignment assigns both $x_1$ and $x_3$ in the third clause to False, resulting in two literals that can be used to satisfy this clause.
$R$ is analogous to the output activation tensor in a DL layer.

\subsubsection{Satisfiability Matrix}
By computing the column-wise minimum of the Result Matrix $R$, we derive the Satisfiability Matrix $S\in \mathbb{Z}_{\geq 0}^{1 \times N}$.
Each entry in $S$ represents the minimum number of literals satisfied by the corresponding assignment in any clause and indicates the satisfiability of each assignment. 
If the per-column minimum is zero, the corresponding assignment is not able to satisfy all the clauses because there exist at least one clause in which the assignment fails to satisfy even a single literal. 
If the per-column minimum is nonzero, the corresponding assignment is able to satisfy all the clauses because the assignment satisfies at least one literal in every clause of the problem. 
If the maximum value of the Satisfiability Matrix $S$ is nonzero, at least one out of the $N$ assignments is a satisfying assignment, and the problem is satisfiable. 
However, if the maximum value of $S$ is zero, there is no satisfying assignments among those attempted, and the problem is indeterminate. 
Note that we can declare UNSAT only if all possible assignments have been exhausted.
In this paper, we focus only on satisfying assignments, and leave UNSAT problems to future work.
%For now, we focus on optimizing for satisfying assignments.

\subsection{Differentiable Solving}
\label{sec:training}
\begin{figure}[t]
    \centering
    \includegraphics[trim=0pt 0pt -20pt 0pt, clip, angle=-90, width=0.83\linewidth]{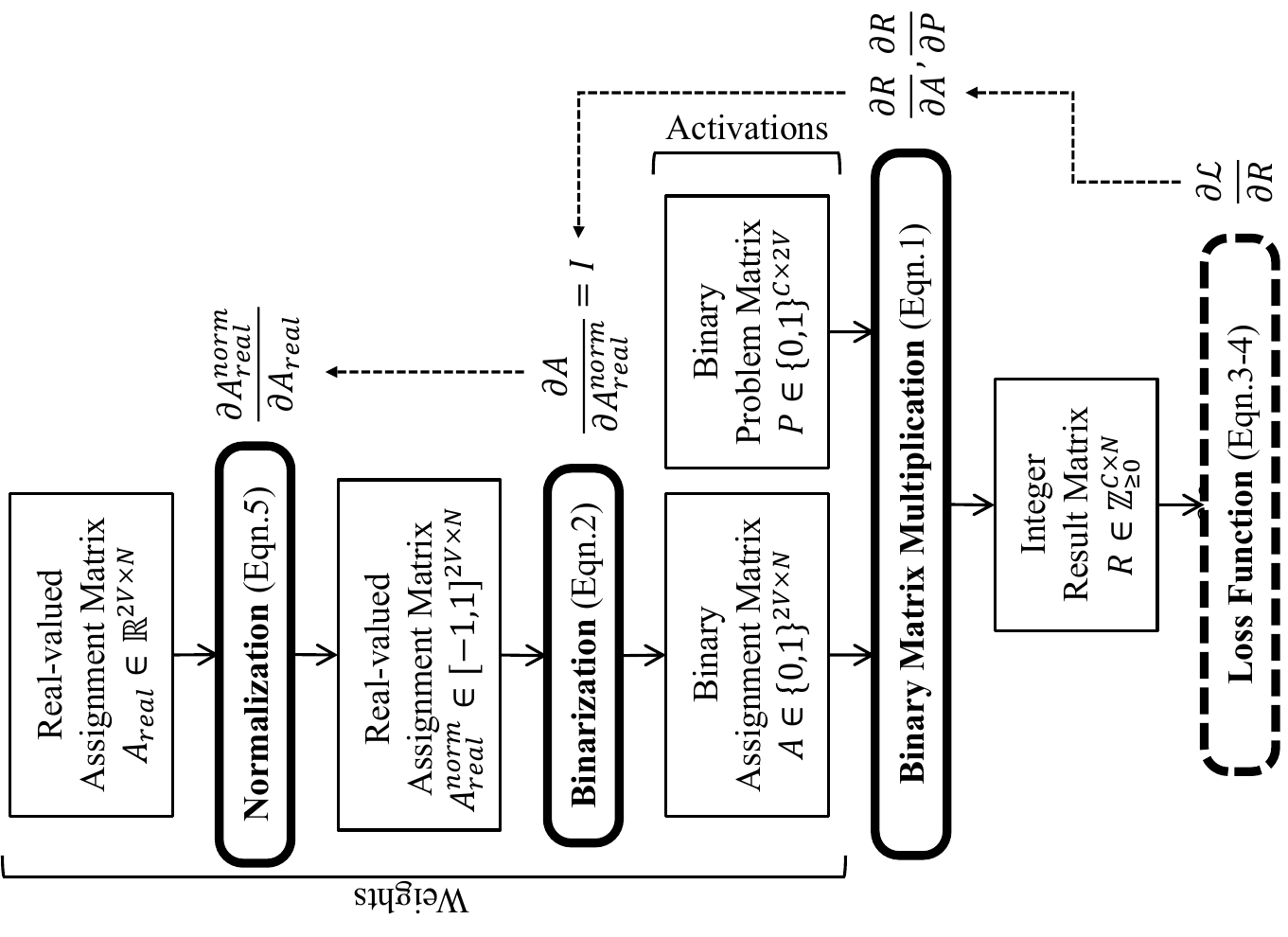}
    \caption{
    \textbf{Computation graph of the binary matrix multiplication for SAT solving} --
    %Forward propagation (solid arrows) and backward propagation (dash arrows) to and from the loss function, respectively, are illustrated.
    In the forward pass (solid arrows), the trainable weights are normalized and binarized prior to multiplication.
    In the backward pass (dash arrows), an identity function is used as a proxy for gradient propagation through the non-differentiable Binarization layer.
    }
    \label{fig:graph}
\end{figure}

We leverage forward and backward passes to perform differentiable SAT solving based on our formulation in Section~\ref{sec:encoding}.
For convenience, we can interpret the matrix multiplication $R=PA$ defined in Section~\ref{sec:encoding} as a Linear layer, where $P$ serves as the non-trainable input activation tensor and $A$ represents the trainable weight parameter tensor.
The computation graph of the binarized matrix multiplication layer used for SAT solving is detailed in Figure~\ref{fig:graph}.
During differentiable optimization, the forward pass (illustrated by solid arrows) evaluates the current set of assignments on the SAT clauses by computing the following matrix multiplication.
\begin{equation}
\label{eqn:matmul}
    R=PA
\end{equation}
Subsequently, the backward pass (illustrated by dash arrows) computes the gradients in respect to the assignment ${\partial \mathcal{L}}/{\partial A}$ and updates the assignments $A$ accordingly to maximize satisfiability. Here $\mathcal{L}$ denotes the loss function that encodes satisfiability.
Because the activation tensor is not trainable in a Linear layer, we can use the 0/1 Problem Matrix $P$ directly as the input that is repeatedly passed into the matrix multiplication at each iteration.
However, to enable proper flow of gradients in the backward pass, we use a real-valued matrix $A_{real}\in \mathbb{R}^{2V \times N}$ in place of the binary matrix $A$ as the trainable weight tensor of the matrix multiplication under the hood.
The binary Assignment Matrix $A$ defined in Section~\ref{sec:encoding} is simply the binarized version of $A_{real}$ based on this binarization function $B$.
\begin{equation}
\label{eqn:binarization}
A=B(A_{real})=\operatorname{clip}\left(\operatorname{sign} (A_{real}),0,1\right)
\end{equation}

%\subsubsection{Optimization Objective}
The training objective for SAT solving aims to achieve satisfiability, meaning that the Satisfiability Matrix $S$ contains at least one nonzero value.
As a proxy, our differentiable objective function is formulated to maximize the sum of the Satisfiability Matrix $S$, giving the loss function $\mathcal{L}$ as follows.
\begin{equation}
\label{eqn:loss}
\mathcal{L} = -\sum_{i=1}^{N} S_i
\end{equation}
However, $S$ is in turn derived from the per-column minimums of the Result Matrix $R$, where the minimum function is not differentiable.
Therefore, we use a differentiable smooth minimum function instead of a hard minimum on each column of $R$, denoted $R_i$, to facilitate gradient propagation.
By applying the smooth minimum function, the Satisfiability Matrix $S$ in the loss function $\mathcal{L}$ is defined such that
\begin{equation}
\label{eqn:smoothmin}
S_i=\mathrm{SmoothMin}(R_i) = \frac{\sum_{j=1}^C R_{ij} e^{-\tau R_{ij}}}{\sum_{j=1}^C e^{-\tau R_{ij}}}
\end{equation}
where $\mathrm{SmoothMin}$ converges to a minimum function as $\tau$ approaches $+\infty$.
For the Binarization Layer in Figure~\ref{fig:graph}, we binarize the parameter tensor $A$ using the $B(x)$ function in Equation~\ref{eqn:binarization} by clamping all positive values to 1 and all non-positive values to 0. 
Following the convention of quantization-aware training~\cite{jacob2018quantization}, we don't explicitly compute the gradients of the Binarization Layer, but use straight-through estimator (STE) instead to pass gradients as-is~\cite{yin2019understanding}.

%% file: 4_system.tex
\section{Gradient-Guided Hybrid Solving}
\label{sec:system}

\begin{figure}
    \centering
    \includegraphics[trim=8pt 225pt 660pt 2pt, clip, width=0.95\linewidth]{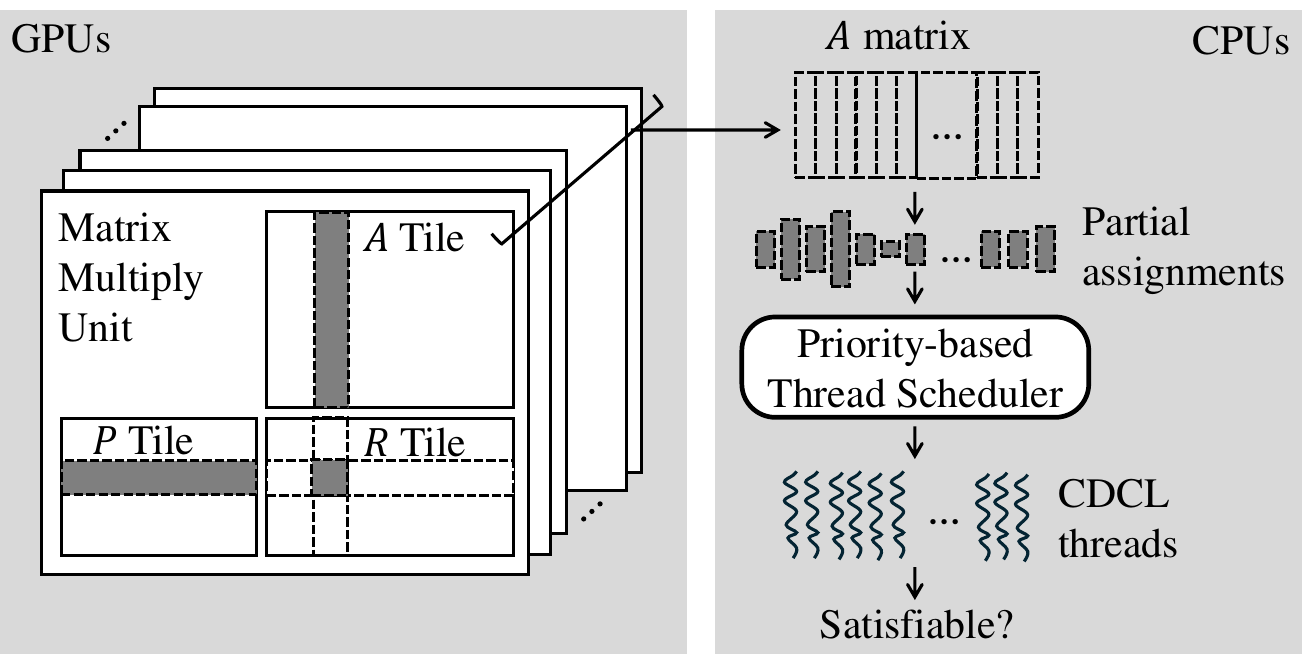}
    \caption{\textbf{Hybrid SAT solving on GPU-CPU system} -- GPUs perform differentiable optimization on matrix-multiply units. CPUs leverage CDCL search on many threads to determine complete satisfying assignments.}
    \label{fig:system}
\end{figure}

Building on the differentiable SAT formulation from Section~\ref{sec:formulation}, we propose a hybrid solving architecture that synergistically combines GPU-driven differentiable optimization with CPU-based combinatorial search, leveraging the complementary strengths of a hybrid GPU-CPU system.
On the GPU, gradient-guided optimization enables rapid exploration of the assignment space through massively parallel matrix operations to evaluate many assignment candidates simultaneously.
On the CPU, CDCL refines promising candidates from the GPU using heuristic-guided conflict-driven search. 
%In essence, the GPU focuses on broad, parallel exploration, while the CPU specializes in precise, sequential refinement, exploiting their respective computational advantages.
Exploiting their respective computational advantages, the GPU focuses on broad, parallel exploration to identify high-likelihood subspaces, while the CPU specializes in precise, sequential refinement to resolve remaining conflicts in narrowed subspaces.
%Using task-specific specialization, the GPU is tasked with identifying high-likelihood subspaces, whereas the CPU is tasked with resolving remaining conflicts in narrowed subspaces.
This approach bridges the gap between gradient-based optimization and conflict-driven search techniques.
The hybrid SAT solving system is shown in Figure~\ref{fig:system}.
%Figure~\ref{fig:system} shows the hybrid SAT solving system.

\subsection{Gradient-based Optimization on GPU}
\label{sec:grad}

Based on the matrix encoding from Section~\ref{sec:encoding}, we construct the Problem Matrix $P$ by parsing each clause in the CNF of the problem, followed by initializing the $N$ assignments in the real-valued Assignment Matrix $A_{real}$ with random values from the standard normal distribution.
This initialization ensures that binarizing 
$A_{real}$ to $A$ yields a balanced number of 0's and 1's, equivalent to assigning approximately half the variables to True and False, respectively.
Once the problem is constructed in the form of our $R=PA$ layer, we iteratively optimize $A_{real}$, and in turn the binary assignments in $A$, by performing forward and backward passes as described in Section~\ref{sec:training}.

For the optimization process, we use the AdamW optimizer with an initial learning rate of $10^{-1}$ and a final learning rate of $10^{-15}$.
We follow a stepwise decay schedule that reduces the learning rate by a factor of 10 every 30 iterations.
This schedule enables aggressive early exploration (using high learning rate) followed by more careful refinement (using small learning rate), which can effectively reduce the number of remaining UNSAT clauses.
We perform periodic restart by resetting the learning rate back to $10^{-1}$ after every 360 iterations.
Our learning rate reset strategy helps navigate around previously best assignments (i.e., local minima) in a rugged optimization landscape and converge towards superior assignments.
%We find empirically that these settings perform well across problems of varying sizes and characteristics. 
%Accordingly, we apply them consistently throughout all experiments reported in Section~\ref{sec:experiments}.
We apply these settings consistently on all experiments reported in Section~\ref{sec:experiments}.

To further improve the efficiency of learning the best variable assignments, we perform per-variable normalization on the real-valued Assignment Matrix $A_{real}$ to enable information sharing on a per-variable basis among all the attempted assignments.
% being attempted.
For row $j$ in $A_{real}$, we compute the mean of the row and scale each element in the row by the mean as shown in Equation~\ref{eqn:normalization}.
% \begin{equation}
% \label{eqn:mean}
% \mu_j=\frac{1}{N}\sum_{i=1}^{N} A_{real, ij}
% \end{equation}
% \begin{equation}
% \label{eqn:normalization}
% A_{real,ij}^{norm}=\frac{A_{real,ij}}{\mu_j}
% \end{equation}
\begin{equation}
\label{eqn:normalization}
A_{real,ij}^{norm}=\frac{A_{real,ij}}{\frac{1}{N}\sum_{i=1}^{N} A_{real, ij}}
\end{equation}
We append this normalization operation in front of the Binarization Layer, as shown in Figure~\ref{fig:graph}.
It is fully differentiable and able to propagate gradients in the standard fashion.
%Mirroring techniques in BNN training, per-variable normalization stabilizes training and accelerates convergence by balancing variable scales~\cite{huang2023normalization}.

Lastly, we observe that our Problem Matrix $P$ is highly sparse.
Therefore, we encode $P$ using the compressed sparse format, enabling highly efficient sparse matrix multiplications in both the forward and backward passes. 
This encoding results in dramatic reduction in memory usage and significant acceleration in differentiable optimization, helping scale our technique to very large problems.

% At each learning rate restart, we also perform per-variable assignment mixing to influence subsequent exploration. 
% The highest-performing assignment $A_{i^*}$ (with the maximum number of satisfied clauses) is identified, and its real-valued weights $A_{real,i^*}$ are averaged into those of inferior candidates as shown in Equation~\ref{eqn:mixing}.
% \begin{equation}
% \label{eqn:mixing}
% A_{real,i}^{mixed}=\frac{1}{2} (A_{real,i}+A_{real,i^*}), i\neq i^*
% \end{equation}
% These blended weights initialize subsequent training iterations, effectively transferring knowledge from high-quality solutions to lower-quality solutions to better guide the exploration.
% Our assignment mixing strategy shares a core principle with Model Soup's parameter mixing technique which has demonstrated success in improving accuracy and robustness in model training~\cite{pmlr-v162-wortsman22a}.
% After mixing the assignments, we reset the optimizer with the newly mixed Assignment Matrix $A_{real}^{mixed}$ to align the optimizer's internal states with the updated parameter values.

% Additional optimization strategies
% Shuffling, diversity
%coordinate descent

\subsection{CDCL-based Refinement on CPU}

Once the optimization on the GPU converges, approximated by satisfying over 99\% of the clauses in the problem, we extract the truth values of the $k$ most confident variables from each candidate assignment, where $k$ is 0.01\% of the number of variables but at least 20.
Confidence is quantified using the final gradient magnitudes of the variables as a proxy, where smaller absolute gradient values indicate variables who have higher confidence in their current assignments.
The set of high-confidence variables (with minimal gradient magnitudes) along with their respective assignments are selected to initialize the CDCL solver, as shown in Figure~\ref{fig:system}.

On the CPU side, each thread receives a partial initialization from the GPU and executes a CDCL-based SAT solver instance from the partial initialization, leveraging conflict-driven learning to resolve the remaining variables. 
The partial initialization is meant to assist the CDCL solvers in deriving satisfying solutions in significantly less time than starting from scratch.
The number of partially initialized parallel CDCL instances scales with the number of CPU threads available on the system, akin to a portfolio solver strategy.
However, we distinguish from prior portfolio solvers by starting with confident partial assignments from GPU-driven gradient-based optimization instead of starting from scratch from random seeds and/or using different solver configurations.
If the number of available CPU threads is insufficient to dispatch the entirety of the assignments in the Assignment Matrix $A$, we prioritize the assignments with higher number of satisfied clauses. 
%The number of satisfied clauses for each candidate assignment can be conveniently extracted from the Result Matrix $R$ by thresholding and summation.

%Empirically, we partially initialize just the most confident 0.01\% of variables with a minimum of 20 variables.
Anchoring just a small fraction of variables may seem insignificant at first sight; but it can dramatically improve solver efficiency in multiple aspects.
%While this may seem insignificant at first sight, it can dramatically improve solver efficiency in multiple aspects.
Theoretically, assigning $V^*$ variables prunes the search space exponentially by $2^{V^*}$, which represents orders-of-magnitude reduction that effectively narrows the CDCL solver’s focus to high-likelihood subspaces~\cite{davis1962machine}.
Practically, these assignments trigger the BCP procedure which simplifies clauses by removing satisfied literals and deduces additional variable assignments through unit propagation. 
%Moreover, assigned variables are less likely to induce conflicts, allowing CDCL to focus on resolving ambiguities in the remaining problem structure.
Industrial SAT instances, typically characterized by a small subset of variables influencing numerous clauses simultaneously, amplify the aforementioned benefits by enabling cascading simplification and conflict resolution opportunities~\cite{ansotegui2009structure}.
Essentially, our approach transforms CDCL from under-informed search into targeted refinement, achieving efficiency gains atypical through random or heuristic initialization alone.
Without gradient guidance, randomly initializing the same number or same subset of variables would provide minimal benefit to CDCL.%, as shown in Section~\ref{sec:exp_cdcl}.

% Algorithm box?

%% file: 5_experiments.tex
\section{Experiments}
\label{sec:experiments}

We prototype our hybrid solver on an NVIDIA DGX GB200 accelerated computing system instance equipped with four NVIDIA Blackwell B200 GPUs and two NVIDIA Grace CPUs each with 72 ARM Neoverse V2 cores~\cite{nvidia2025dgxgb200}.
Although there are four GPUs and 144 CPU cores, we limit our solver to use up to only two GPUs and 141 CPU threads. %to minimize excessive resource usage and potential contention.
%, and system instabilit.
Our solver leverages the PySAT framework~\cite{imms-sat18} for parsing input problems and interfacing with CDCL SAT solvers.
Gradient-based optimization is executed on the GPU using PyTorch 2.2~\cite{paszke2019pytorch}, while CDCL-based refinement is run on the CPU with the CDCL solver \mbox{CaDiCaL 1.9.5} ~\cite{cadical195}.
%Each CDCL thread operates as a single independent \mbox{CaDiCaL} process, and each \mbox{CaDiCaL} instance runs in a single thread.
Our setup mimics the system shown in Figure~\ref{fig:system}.
To ensure fair comparison, we use the same \mbox{CaDiCaL} solver for our baseline experiments.
\mbox{CaDiCaL} is a state-of-the art SAT solver used in industry and one of the top contenders in the SAT Competition.
We evaluate runtime performance on all satisfiable benchmark problems from the 2024 SAT Competition~\cite{heule2024proceedings}.

\input{t5_speedup}

%In Table~\ref{tab:speedup}, we present the runtime results of our hybrid solver running on the DGX GB200 system by benchmarking its performance against pure-CDCL single-threaded \mbox{CaDiCaL} running on the CPU in the same system.
In Table~\ref{tab:speedup}, we first detail the runtime results of our hybrid solver on selected representative problems of various sizes and types and compare them to the results of the baseline solver \mbox{CaDiCaL}.
We would like to point out that our solver is always at least as competitive as \mbox{CaDiCaL}, as we dedicate one thread to running \mbox{CaDiCaL} from scratch. %without any partial initialization; thus, our solver can solve any instance that \mbox{CaDiCaL} can solve.
%In addition, we de-emphasize problems that CDCL can solve efficiently and quickly without additional guidance.
%Benchmarks listed in Table~\ref{tab:speedup} span a range of 180 to 30,587 variables and 432 to 192,064 clauses, with clause-to-variable ratios from 2.4 to 64.98.
For each benchmark, alongside basic problem statistics, we list the runtime in seconds for both \mbox{CaDiCaL} and our solver.
Due to the hybrid nature of our solver, we report the total runtime (denoted \texttt{Total}), as well as the runtime spent in the gradient-guided optimization phase (denoted \texttt{Gradient}) and the CDCL-based refinement phase (denoted \texttt{CDCL}).
To quantify the performance gain, we include the total speedup relative to \mbox{CaDiCaL}, as well as the speedup achieved during the CDCL refinement phase alone.
Because our hybrid solver internally uses the same \mbox{CaDiCaL} solver as our baseline \mbox{CaDiCaL}, the latter speedup precisely reflects the benefit of initializing with gradient-guided partial assignments.

\begin{figure}[t]
    \centering
    \includegraphics[trim=660pt 445pt 0pt 20pt, clip, width=0.8\linewidth]{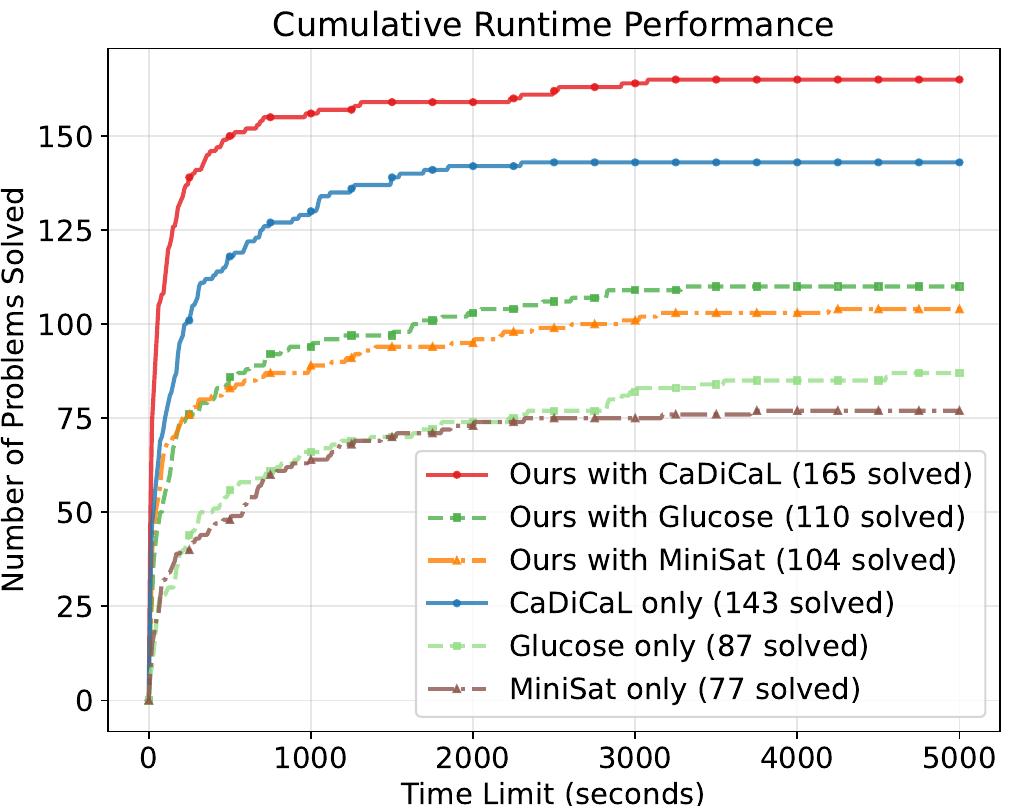}
    \caption{\textbf{Cumulative performance curve} -- Compares the number of solved problems for a given time limit between our solvers and baseline solvers.}
    \label{fig:cumulative}
\end{figure}

Benchmarks in Table~\ref{tab:speedup} are sorted by their total speedup.
Many problems achieving significant speedup have been omitted in Table~\ref{tab:speedup} due to space constraints, but are included in the cumulative runtime performance plot in Figure~\ref{fig:cumulative}.
In Table~\ref{tab:speedup}, benchmarks for which \mbox{CaDiCaL} times out after two hours are listed first.
%All runtimes are averaged among two or more runs.
Compared to \mbox{CaDiCaL}, our solver achieves a total runtime speedup of 5.03x to 258.08x on the listed benchmarks.
Considering CDCL alone, our partial initialization strategy observes a speedup from 5.94x to 1367.73x.
Notably, our solver sees greater than 10x total speedup for benchmarks with up to more than 74,000 variables and 393,000 clauses.
Our solver also solves many problems within 1000 seconds for which \mbox{CaDiCaL} times out even after two hours.

Specifically, our solver achieves greater than 100x speedup for the two \texttt{x9} benchmarks, solving them in seconds compared to more than 20 minutes spent by \mbox{CaDiCaL}.
Our solver also achieves greater than 20x speedup for the \texttt{j3037} problem with a clause-to-variable ratio of 4.53, close to the critical 4.25 threshold where a problem tends to be hardest to solve~\cite{KirkpatrickSelman94}. 
Furthermore, we see a speedup of 26.31x on the \texttt{ex065\_25} problem with 74,776 variables and 393,322 clauses, representing problems of non-trivial size.
%Our solver is also able to handle \texttt{mp1-klieber} which contains 30,587 variables and 91,703 clauses.
%Finally, it is important to note that there are many problems (listed at the top of Table~\ref{tab:speedup}) that our solver solves with non-trivial speedup for which \mbox{CaDiCaL} times out even after two hours.
On the problems for which \mbox{CaDiCaL} times out, our solver solves \texttt{6g\_6color} within a minute even though it contains 77,328 variables and 8,485,566 clauses.
Our solver also solves \texttt{Nb13T165} in 22 minutes while handling over one million variables and five million clauses.
Considering all 179 satisfiable problems in the Competition, our solver achieves an average total speedup of 27.30x.% based on SAT Competition rules. 
%For these benchmarks, we cannot determine their actual speedup due to unknown baseline \mbox{CaDiCaL} runtime.
%Based on \mbox{MiniSat's} two-hour timeout, the speedup is at least on the order of an hour, but the actual speedup coupld be much greater.
%For these benchmarks, we report the speedup based on the two-hour timeout we set for MiniSat.
%The actual speedups could be much greater than reported in Table~\ref{tab:speedup}.

Figure~\ref{fig:cumulative} demonstrates the cumulative runtime performance on the entire satisfiable problem set.
We show the performance curves for our solver along with baselines and variants.
%compared to both \mbox{CaDiCaL} and \mbox{MiniSat}.
We include \mbox{MiniSat} as a reference because it is the foundational SAT solver on which others are developed.
We also include \mbox{Glucose}~\cite{AudemardSimonIJCAI09} for its different solving heuristics.
Each point on a curve denotes the number of problems solved within the given time limit.
The plot demonstrates the advantage of our solver over the baselines. 
Our solver, integrated with either \mbox{CaDiCaL}, \mbox{Glucose}, or \mbox{MiniSat}, solves more problems within shorter time limits compared to using each respective solver alone.
This shows that gradient guidance is effective for different CDCL solvers.
In particular, our solver with \mbox{CaDiCaL} achieves the highest performance, solving a total of 165 problems within 5000 seconds. 
That is 22 more solved than \mbox{CaDiCaL} alone.

Figure~\ref{fig:cumulative} also highlights the strength our solver at different time limits.
Up to 250 seconds, the curves for our solver climb rapidly regardless of the underlying CDCL solver, indicating that our approach solves problems much more efficiently than baselines.
As the time limit increases towards 1000 seconds, the performance gap remains pronounced.
Evident in the slopes of the curves, our solver with \mbox{CaDiCaL} outpaces all other methods shown by solving more problems within the same given time.
Beyond 2000 seconds, our solver continues to improve while \mbox{CaDiCaL}'s performance plateaus.
Overall, our solver with \mbox{CaDiCaL} achieves a PAR2 SAT score~\cite{heule2024proceedings} of 972.28 compared to 2228.22 for \mbox{CaDiCaL}.
We attain this score without clause sharing or other parallel solving techniques.

%% file: t5_speedup.tex
\begin{table*}[tbh]
\setlength{\tabcolsep}{3.5pt}
\centering
\caption{
\textbf{Selected runtime comparison} -- 
\texttt{Total} denotes the total runtime of our solver. 
\texttt{Gradient} denotes the gradient-guided optimization portion. 
\texttt{CDCL} denotes the CDCL portion.
Timeout is 7200 seconds.
Speedup compares our solver against \mbox{CaDiCal}.
}
\label{tab:speedup}
%\begin{tabular}{|l|S[table-format=5.0]|S[table-format=6.0]|S[table-format=2.2]|cccc|cc|}
\begin{tabular}{|l|r|r|r|rrrr|rr|}
\hline
\multicolumn{1}{|c|}{\multirow{3}{*}{\textbf{Benchmark}}} & \multicolumn{1}{c|}{\multirow{3}{*}{\textbf{\#Variables}}} & \multicolumn{1}{c|}{\multirow{3}{*}{\textbf{\#Clauses}}} & \multicolumn{1}{c|}{\multirow{3}{*}{\textbf{Ratio}}} & \multicolumn{4}{c|}{\textbf{Runtime (seconds)}}                                                                                                                            & \multicolumn{2}{c|}{\textbf{Speedup}}                                                      \\ \cline{5-10} 
\multicolumn{1}{|r|}{}                                    & \multicolumn{1}{r|}{}                                      & \multicolumn{1}{r|}{}                                    & \multicolumn{1}{r|}{}                                & \multicolumn{1}{c|}{\multirow{2}{*}{\textbf{CaDiCaL}}} & \multicolumn{3}{c|}{\textbf{Ours}}                                                                                & \multicolumn{1}{c|}{\multirow{2}{*}{\textbf{CDCL}}} & \multicolumn{1}{c|}{\multirow{2}{*}{\textbf{Total}}} \\ \cline{6-8}
\multicolumn{1}{|r|}{}                                    & \multicolumn{1}{r|}{}                                      & \multicolumn{1}{r|}{}                                    & \multicolumn{1}{r|}{}                                & \multicolumn{1}{r|}{}                                  & \multicolumn{1}{c|}{\textbf{Gradient}} & \multicolumn{1}{c|}{\textbf{CDCL}} & \multicolumn{1}{c|}{\textbf{Total}} & \multicolumn{1}{c|}{}                               & \multicolumn{1}{r|}{}                                \\ \hline

pcmax-scheduling-m12-8049-55035-SAT & 8048 & 55035 & 6.84 & \multicolumn{1}{r|}{Timeout} & \multicolumn{1}{r|}{5.19} & \multicolumn{1}{r|}{30.76} & \multicolumn{1}{r|}{35.95} & \multicolumn{1}{r|}{Unknown} & Unknown\\ \hline
6g\_6color\_366\_050\_04 & 77328 & 8485566 & 109.73 & \multicolumn{1}{r|}{Timeout} & \multicolumn{1}{r|}{47.44} & \multicolumn{1}{r|}{3.50} & \multicolumn{1}{r|}{50.94} & \multicolumn{1}{r|}{Unknown} & Unknown\\ \hline
002 & 4128 & 126564 & 30.66 & \multicolumn{1}{r|}{Timeout} & \multicolumn{1}{r|}{4.93} & \multicolumn{1}{r|}{102.27} & \multicolumn{1}{r|}{107.20} & \multicolumn{1}{r|}{Unknown} & Unknown\\ \hline
pcmax-scheduling-m40-26287-324155-SAT & 26286 & 324155 & 12.33 & \multicolumn{1}{r|}{Timeout} & \multicolumn{1}{r|}{7.52} & \multicolumn{1}{r|}{99.86} & \multicolumn{1}{r|}{107.38} & \multicolumn{1}{r|}{Unknown} & Unknown\\ \hline
Folkman-185-19924337 & 17019 & 19874 & 1.17 & \multicolumn{1}{r|}{Timeout} & \multicolumn{1}{r|}{4.53} & \multicolumn{1}{r|}{281.05} & \multicolumn{1}{r|}{285.58} & \multicolumn{1}{r|}{Unknown} & Unknown\\ \hline
mdp-32-11-sat & 1038 & 5842 & 5.62 & \multicolumn{1}{r|}{Timeout} & \multicolumn{1}{r|}{9.18} & \multicolumn{1}{r|}{659.29} & \multicolumn{1}{r|}{668.47} & \multicolumn{1}{r|}{Unknown} & Unknown\\ \hline
Break\_20\_72.xml & 53844 & 270142 & 5.01 & \multicolumn{1}{r|}{Timeout} & \multicolumn{1}{r|}{9.28} & \multicolumn{1}{r|}{956.74} & \multicolumn{1}{r|}{966.02} & \multicolumn{1}{r|}{Unknown} & Unknown\\ \hline
Nb13T165 & 1389990 & 5543200 & 3.98 & \multicolumn{1}{r|}{Timeout} & \multicolumn{1}{r|}{58.06} & \multicolumn{1}{r|}{1251.36} & \multicolumn{1}{r|}{1309.42} & \multicolumn{1}{r|}{Unknown} & Unknown\\ \hline
fermat-931960058139995587 & 12257 & 69473 & 5.66 & \multicolumn{1}{r|}{Timeout} & \multicolumn{1}{r|}{6.86} & \multicolumn{1}{r|}{2483.75} & \multicolumn{1}{r|}{2490.61} & \multicolumn{1}{r|}{Unknown} & Unknown\\ \hline
1-ET-512-K-96.sanitized & 45616 & 2068700 & 45.35 & \multicolumn{1}{r|}{Timeout} & \multicolumn{1}{r|}{85.93} & \multicolumn{1}{r|}{2435.51} & \multicolumn{1}{r|}{2521.44} & \multicolumn{1}{r|}{Unknown} & Unknown\\ \hline
apn-sbox5-cut3-symmbreak & 21240 & 86081 & 4.05 & \multicolumn{1}{r|}{Timeout} & \multicolumn{1}{r|}{4.67} & \multicolumn{1}{r|}{2913.15} & \multicolumn{1}{r|}{2917.82} & \multicolumn{1}{r|}{Unknown} & Unknown\\ \hline

x9-11053.sat.sanitized & 550 & 4951 & 7.40 & \multicolumn{1}{r|}{1244.63} & \multicolumn{1}{r|}{3.91} & \multicolumn{1}{r|}{0.91} & \multicolumn{1}{r|}{4.82} & \multicolumn{1}{r|}{1367.73x} & 258.08x\\ \hline
x9-12014.sat.sanitized & 600 & 5405 & 9.01 & \multicolumn{1}{r|}{1692.45} & \multicolumn{1}{r|}{3.95} & \multicolumn{1}{r|}{5.53} & \multicolumn{1}{r|}{9.48} & \multicolumn{1}{r|}{306.05x} & 178.50x\\ \hline
%j3045\_10\_rggt\_b & 14751 & 63197 & 4.28 & \multicolumn{1}{r|}{304.77} & \multicolumn{1}{r|}{4.89} & \multicolumn{1}{r|}{36.12} & \multicolumn{1}{r|}{41.01} & \multicolumn{1}{r|}{8.43} & 7.43\\ \hline
combined-crypto1-wff-seed-1-wffvars-450 & 1729 & 21718 & 12.56 & \multicolumn{1}{r|}{176.26} & \multicolumn{1}{r|}{4.59} & \multicolumn{1}{r|}{1.24} & \multicolumn{1}{r|}{5.83} & \multicolumn{1}{r|}{142.15x} & 30.21x\\ \hline
j3037\_9\_rggt\_b & 18913 & 85707 & 4.53 & \multicolumn{1}{r|}{396.37} & \multicolumn{1}{r|}{4.82} & \multicolumn{1}{r|}{10.11} & \multicolumn{1}{r|}{14.93} & \multicolumn{1}{r|}{39.21x} & 26.55x\\ \hline
ex065\_25 & 74776 & 393322 & 5.26 & \multicolumn{1}{r|}{1492.35} & \multicolumn{1}{r|}{10.35} & \multicolumn{1}{r|}{46.37} & \multicolumn{1}{r|}{56.72} & \multicolumn{1}{r|}{32.18x} & 26.31x\\ \hline
rbsat-v760c43649gyes3 & 760 & 43639 & 57.42 & \multicolumn{1}{r|}{129.96} & \multicolumn{1}{r|}{4.39} & \multicolumn{1}{r|}{1.08} & \multicolumn{1}{r|}{5.47} & \multicolumn{1}{r|}{120.33x} & 23.75x\\ \hline
Circuit\_multiplier22 & 1013 & 18793 & 18.55 & \multicolumn{1}{r|}{233.56} & \multicolumn{1}{r|}{5.31} & \multicolumn{1}{r|}{7.42} & \multicolumn{1}{r|}{12.73} & \multicolumn{1}{r|}{31.48x} & 18.35x\\ \hline
004 & 4288 & 132576 & 30.92 & \multicolumn{1}{r|}{1490.67} & \multicolumn{1}{r|}{5.04} & \multicolumn{1}{r|}{95.94} & \multicolumn{1}{r|}{100.98} & \multicolumn{1}{r|}{15.53x} & 14.76x\\ \hline
af-synthesis\_stb\_50\_40\_9\_sat & 15981 & 124882 & 7.81 & \multicolumn{1}{r|}{740.29} & \multicolumn{1}{r|}{5.75} & \multicolumn{1}{r|}{48.07} & \multicolumn{1}{r|}{53.82} & \multicolumn{1}{r|}{15.40x} & 13.75x\\ \hline
noL-11-20.sanitized & 1419 & 7841 & 5.52 & \multicolumn{1}{r|}{82.25} & \multicolumn{1}{r|}{3.91} & \multicolumn{1}{r|}{5.04} & \multicolumn{1}{r|}{8.95} & \multicolumn{1}{r|}{16.32x} & 9.19x\\ \hline
preimage\_80r\_495m\_160h\_seed\_379 & 56108 & 224206 & 3.99 & \multicolumn{1}{r|}{216.86} & \multicolumn{1}{r|}{6.61} & \multicolumn{1}{r|}{36.52} & \multicolumn{1}{r|}{43.13} & \multicolumn{1}{r|}{5.94x} & 5.03x\\ \hline
%qwh.60.1728.shuffled-as.sat03-1659 & 25162 & 542447 & 21.56 & \multicolumn{1}{r|}{67.81} & \multicolumn{1}{r|}{7.47} & \multicolumn{1}{r|}{9.36} & \multicolumn{1}{r|}{16.83} & \multicolumn{1}{r|}{7.24} & 4.03\\ \hline
\hline
\textbf{All satisfiable problems} & - & - & - & \multicolumn{1}{r|}{-} & \multicolumn{1}{r|}{-} & \multicolumn{1}{r|}{-} & \multicolumn{1}{r|}{-} & \multicolumn{1}{r|}{\textbf{Average}} & \textbf{27.30x}\\ \hline

\end{tabular}
\end{table*}

%% file: 6_conclusions.tex
\vspace{8pt}
\section{Conclusions}
\label{sec:conclusions}
The integration of GPU-accelerated gradient-driven optimization with CPU-based conflict-driven search represents a promising frontier for scalable SAT solving. 
By reformulating SAT as a differentiable binarized matrix multiplication task, our hybrid approach effectively balances the strengths of massively parallel exploration on GPU with targeted sequential search on CPU.
%The GPU leverages neural network training to rapidly evaluate SAT clauses and identify promising subspaces by optimizing assignments with gradients.
%Many CPU threads build upon partial assignments from these subspaces by employing sequential search to further traverse targeted subspaces and derive complete assignments.
By demonstrating its benefits on a commercial GPU-CPU hybrid system, our results highlight the potential of combining modern differentiable optimization with classical search algorithms to overcome inherent parallelism limitations in logical reasoning and combinatorial optimization. 